\documentclass[preprint,showkeys,eqsecnum,aps,floatfix,amsmath,amssymb]{revtex4}
\usepackage{graphicx}
\usepackage{dcolumn}
\usepackage{bm}
\begin{document}
\newcommand\lsim{\alt}
\newcommand\stress[1]{{\it#1\/}}
\newcommand\whatis[1]{(\ref{#1})}
\newcommand\ie{{\it i.e.\/}}
\newcommand\etal{{\it et al.\/}}
\def\PD#1#2{\frac{\partial#1}{\partial#2}}

\newcommand\Abar{\bar{A}}
\newcommand\Kbar{\bar{K}}
\newcommand\Sbr{\Sigma_{\beta|\gamma}}
\newcommand\Sab{\Sigma_{\alpha|\beta}}
\newcommand\Sabr{\Sigma_{\alpha|\beta\gamma}}
\newcommand\Sar{\Sigma_{\alpha|\gamma}}
\newcommand\mt{\widetilde m}
\newcommand\htilde{\tilde h}
\newcommand\Sm{S^\pm_M(\mt)}

\newcommand\celc{\hbox{\thinspace$^\circ\hbox{\rm C}$}}
\newcommand\stu{\hbox{\hbox{erg}/\hbox{cm}$^2$}}
\newcommand\kelvin{\hbox{K}}
\newcommand\gcm{\hbox{g}/\hbox{cm}^3}
\newcommand\Ao{\thinspace{\buildrel \kern0.15em\scriptstyle\circ \over A}}
\newcommand\gl{\buildrel > \over {_<}}
\newcommand\rrho{\check \rho}
\newcommand\rh{\check h}
\newcommand\rK{\check K}

\newcommand\Pexpt{P_{\rm expt}}
\newcommand\Qexpt{Q_{\rm expt}}

\newcommand\emspace{\hbox to 1 em {}}

\title{Interfacial tensions near critical endpoints:\\
experimental checks of EdGF theory}

\author{SHUN-YONG ZINN and MICHAEL E. FISHER*}
\affiliation{%
Institute for Physical Science and Technology,\\
University of Maryland, College Park, MD 20742, USA}%
\date{22 March 2005}

\begin{abstract}
Predictions of the extended de Gennes-Fisher local-functional
theory for the universal scaling functions of interfacial tensions near
critical endpoints are compared with
experimental data.  Various observations of the
binary mixture isobutyric acid $+$ water are correlated to facilitate
an analysis of the experiments of Nagarajan,
Webb and Widom who observed the vapor-liquid interfacial tension
as a function of \stress{both} temperature and density.
Antonow's rule is confirmed and, with the aid of previously studied
{\it universal amplitude ratios\/}, the
crucial analytic ``background'' contribution to the surface tension
near the endpoint is estimated.
The residual singular behavior thus uncovered
is consistent with the theoretical scaling predictions and confirms the
expected lack of symmetry in $(T-T_c)$.  A searching test of theory,
however, demands more precise and extensive
experiments; furthermore, the analysis highlights, a previously noted but
surprising, three-fold discrepancy in the magnitude of the surface
tension of isobutyric acid $+$ water relative to other systems.
\end{abstract}


\maketitle

%
%
\section{Introduction}
\noindent
On passing through a critical endpoint, a binary liquid
mixture of, say, species B and C, in
the presence of the vapor~$\alpha$, exhibits phase separation
below~$T_c$ into
two phases~$\beta$ and~$\gamma$ while above $T_c$ there is only
one homogeneous liquid
phase~$\beta\gamma$~\cite{Rowlinson, Widom77, Widom80, NWW}.
The~$\gamma$ phase, rich in C, may be supposed to have the
higher density so that in a gravitational field it lies below the~$\beta$
phase.  In fact
there are binary mixtures which display lower consolute points so that
phase separation occurs
above~$T_c$ and mixing below~$T_c$; but since there are found to be
no basic differences in
criticality, only upper consolute points will be considered explicitly.

Renormalization group theory has recently shown that the
critical behavior realized
at a critical endpoint lies in the same universality class
as on the critical
locus that results when the pressure is increased so that
the vapor phase $\alpha$ is fully suppressed~\cite{Diehl2000, Diehl2001}.
Nevertheless, near a critical
endpoint, new \stress{bulk} and \stress{interfacial} singularities arise
\cite{Fisher90a, Fisher90b}.  These are of particular interest in the
interfacial tensions~$\Sbr$,
$\Sab$, $\Sar$, and $\Sabr$ corresponding to the various interfaces
indicated by the subscripts \cite{Widom77, Widom80, NWW, Fisher90a,
Fisher90b}.  Indeed, except for the first, namely $\Sbr(T)$, all these tensions
are functions \stress{both} of the temperature, $T$, \stress{and} of the
overall density, $\rho$ (or, equally, of the overall composition).

In terms of the standard notation, with
$t \equiv (T-T_c)/T_c$, the behavior of the \stress{critical}
\stress{surface}
\stress{tension} $\Sbr(T)$ that
\stress{vanishes} at the critical endpoint can be described by
\begin{equation}
\Sbr(T) \approx K |t|^\mu, \quad t \to 0-, \quad (h=0),
\label{eq5.1:sbr}
\end{equation}
where $K$ is a \textit{non}universal critical amplitude while
in $d \enspace (\le 4)$ spatial dimensions
the exponent satisfies the scaling relations
\begin{eqnarray}
\mu &=& 2 \beta + \gamma - \nu = 2 - \alpha - \nu, \nonumber\\
    &=& \beta (\delta + 1) - \nu = (d - 1) \nu,
\label{eq5.1:exprel}
\end{eqnarray}
first derived, specifically for the surface tension, by Widom
\cite{Widom65, Widom72}.  The symbol $h$ in parentheses in
\whatis{eq5.1:sbr} denotes the \stress{ordering} \stress{field}
[conjugate to the order parameter $\sim\negthinspace(\rho - \rho_c)$] that is 
introduced in the scaling description \cite{Widom77, Widom80, Fisher90a,
Fisher90b, Zinn99, Zinn03}.  It is an analytic combination of the
temperature $T$ and the chemical potentials, $\mu_B$ and $\mu_C$, or,
e.g., the chemical potential difference, $\Delta \mu = \mu_C - \mu_B$,
and the pressure, $p$, (all these being thermodynamic fields in the
standard sense); furthermore, $h$ is defined so that it vanishes \stress{at} the
\stress{critical} \stress{endpoint}, specified, say, by $(T_c, \rho_c,
p_c)$, and is identically zero on the whole $(\beta, \gamma)$
coexistence surface (or phase boundary) in the $(T, \mu_B, \mu_C)$
thermodynamic space: see, e.g., Fig.~1 of {\bf I} \cite{Zinn03}.

The other surface tensions between the
noncritical vapor or \stress{spectator} \stress{phase} $\alpha$
and the liquid phases $\beta$, $\gamma$, and $\beta\gamma$ share a common
``background'' contribution, $\Sigma_0(T, h)$, which 
varies analytically and does \stress{not} vanish at
criticality.  Thus we can write \cite{Fisher90a, Fisher90b}
\begin{eqnarray}
\Sabr(T) &\approx& K^+ |t|^\mu + \Sigma_0(T, 0), \quad t \to 0+,
    \quad (h = 0),
    \nonumber\\
\Sab(T) &\approx& K^- |t|^\mu + \Sigma_0(T, 0), \quad t \to 0-,
    \quad (h = 0-),
    \label{eq5.1:sabr}\label{eq5.1:sab}
\end{eqnarray}
where $K^+$ and $K^-$ are again nonuniversal amplitudes.  When,
as generally expected \cite{Rowlinson}, the wetting temperature, $T_W$,
lies below $T_c$, the intermediate $\beta$ phase
spreads over the $\alpha | \gamma$ interface;
then Antonow's rule~\cite{Rowlinson, Zinn03} holds, that is,
\begin{equation}
\Sar(T) = \Sab(T) + \Sbr(T), \quad (h=0) \label{eq5.1:antonow}
\end{equation}
which by (\ref{eq5.1:sbr}) and (\ref{eq5.1:sab}) implies
\begin{equation}
\Sar(T) \approx (K + K^-) |t|^\mu + \Sigma_0(T, 0), \quad t\to0-, \quad h =
{0+}.
\label{eq5.1:sar}
\end{equation}

More generally, on the whole thermodynamic surface bounding the $\alpha$ phase,
we have
\begin{equation}
\Sigma(T, h) = \Delta\Sigma(T, h) + \Sigma_0(T, h),
\label{eq5.1:sigma}
\end{equation}
where $\Delta\Sigma(T,h)$ and $\Sigma_0(T, h)$ are the singular
and regular (or analytic) parts of the surface tension, respectively.
Above $T_c$ one has $\Sigma = \Sabr$, while below $T_c$ one has
$\Sigma = \Sar$ when, by convention, $h > 0$
but $\Sigma = \Sab$ when $h < 0$.

According to general scaling principles
the singular part can be written asymptotically as
\begin{equation}
\Delta\Sigma(T,h) \approx K|t|^\mu S_M^\pm[\mt(T, h)],
\quad\quad\quad\quad \mt\equiv M(T,h)/B|t|^\beta,
\label{eq5.1:Spm}
\end{equation}
where the superscripts $+$ and $-$ stand for $t \gl 0$, respectively:
see {\bf I}.
Because in experiments the density
is more readily accessible than the chemical potentials the argument
$\mt$ of the
scaling functions, $S_M^\pm(\bullet)$, has been chosen
proportional to the order
parameter which, in leading order, may be taken as $M \equiv (\rho -
\rho_c)$. The
coefficient $B$ represents the coexistence curve amplitude according
to $M_0 \approx B|t|^\beta$.  The nonuniversal amplitudes~$B$ and~$K$ are
introduced in \whatis{eq5.1:Spm} to make the scaling functions
$S_M^\pm(\bullet)$ and their arguments
dimensionless.  Then the $S_M^\pm(\bullet)$ are expected to be
\stress{universal} with,
by virtue of Antonow's rule~\whatis{eq5.1:antonow},
\begin{equation}
S_M^-(1) - S_M^-(-1) = 1. \label{eq5.1:SMpmnorm}
\end{equation}

Owing to the technical and conceptual difficulties arising from the fluctuations
of capillary waves, the renormalization group approach has not so far
been successful
in calculating the $\Sm$.  Instead, advances have been made with the aid
of phenomenological theories
based on local-functional concepts.  The pioneering studies, both
theoretical and experimental, have been made by
Widom and his co-workers~\cite{Widom77,
Widom80, NWW}.  However, as noted in \cite{Fisher90a,
Fisher90b}, their theory predicts a ``correction term'' varying as
$|t|^\gamma$ which becomes
more singular than the leading $|t|^\mu$ term when $d > 3 - \eta$
as is relevant in real three-dimensional systems.  This
defect was remedied in the EdGF or ``extended de Gennes-Fisher''
theory proposed in~\cite{Fisher90a, Fisher90b} and implemented recently
in {\bf I} \cite{Zinn03} which forms the basis for the present report.
By using an accurate representation of the
equation of state near criticality~\cite{Zinn99},
the EdGF scaling functions~$\Sm$ have been calculated explicitly
and presented in {\bf I}.
On evaluating the scaling functions
in zero field, one obtains estimates for the Fisher-Upton
ratios \cite{Fisher90a, Fisher90b}, namely,
\begin{equation}
P \equiv (K^+ + K^-)/K = 0.137_5\pm0.002,
\quad Q \equiv K^+/K^- = -0.834\pm0.002. \label{eq5.1:PQ}
\end{equation}

These ratios have been studied experimentally by 
Woermann and coworkers~\cite{aKreuser, Mainzer} for the binary fluid
2,6-dimethyl pyridine $+$ water: their
results are discussed in Sec.~II below.

To go further and test the basic theoretical predictions for
the interfacial scaling functions $S_M^\pm(\mt)$ \stress{away}
from the $h = 0$ axis, the notable experimental data of
Nagarajan, Webb, and Widom ({\bf NWW})~\cite{NWW} for mixtures of
isobutyric acid and water offer, to our knowledge, the only
available opportunity.
For the same mixture, Howland,
Wong, and Knobler ({\bf HWK})~\cite{Knobler} measured the critical
surface tension
$\Sbr(T)$ while Greer~\cite{Greer} measured the densities on the coexistence curve.
These two experiments provide what prove to be valuable
consistency checks and calibrations for the {\bf NWW}
data as will be seen below.

It may be noted that there 
are interesting surface-tension experiments on other
quasi-binary and binary mixtures, such as $n$-octadecane and $n$-nonadecane
in ethane~\cite{Pegg85},
2,5-lutidine in water~\cite{Privat}, and 2-butoxyethanol in
water~\cite{Woerman95}.  However, our attention will focus on the
isobutyric acid $+$ water system since only in this case are there
sufficient measurements of the surface tensions
near the critical endpoint to warrant an attempt to extract the
interfacial scaling functions.

Since the surface tension vanishes at the critical point
as $|t|^\mu$ with $\mu \simeq 1.26$, and hence faster than linearly,
the background contribution to the surface tensions
proves highly
significant as anticipated in {\bf I}.  Thus, in analyzing the
experimental data, it is essential to determine the background
$\Sigma_0(T,h)$ carefully.  By virtue of the analyticity of
the background, we may suppose it can be well represented
near the critical endpoint,
$(t,h)=(0,0)$, by an expansion of the form 
\begin{equation}
\Sigma_0(T, h) = \Sigma_c + \Sigma_t t + \Sigma_h h +
	\Sigma_{tt} t^2 + \Sigma_{th} t h + \Sigma_{hh} h^2 + \cdots.
\label{eq5.1:S0}
\end{equation}
Examination of the {\bf NWW} experimental data (see their Fig.~8)
reveals a
definite upward curvature in the plots of $\Sigma$ vs.\ $T$ at
fixed composition; but the theoretical results for the singular part
$\Delta \Sigma (T,0)$
\stress{alone} shows the opposite, downward curvature above $T_c$: see
{\bf I} Fig.~5(a) and the negative value of $Q$ in \whatis{eq5.1:PQ}.
To capture this dominant behavior, the
expansion~\whatis{eq5.1:S0} must contain at least quadratic terms in $t$.

In the analysis reported by {\bf NWW} of their own data,
the background was assumed to be representable by the form
\begin{equation}
\Sigma_b(T, M) \simeq \Sigma_c + \Sigma_t t + \Sigma_{tM}\thinspace t M.
\label{eq5.1:NWWS0}
\end{equation}
However, this expression is \stress{not} smooth in $t$ and $h$, as expected
on general grounds,
since on the critical
isotherm it implies $(\partial \Sigma_b / \partial t) \simeq \Sigma_t + 
\Sigma_{tM} M$ while $M$ varies as $\hbox{sgn}(h) |h|^{1/\delta}$
with $1/\delta \simeq 0.21$ which is highly singular.  Furthermore, the
{\bf NWW} background assumption
seems to have led to the conclusion that $S_M^+(\mt)$ \stress{above} $T_c$
was indistinguishable from $S_M^-(\mt)$ \stress{below} $T_c$: see {\bf
NWW} Fig.~10.
However, when a suitable background of the form \whatis{eq5.1:S0} is used
in the analysis, one sees
that the data support a difference between $S_M^+(\mt)$ and $S_M^-(\mt)$
as, indeed, predicted theoretically in {\bf I} and expected quite
generally: see Sec.~V below.

\begin{center}
{\bf Units and Exponents}
\end{center}

\noindent
For convenience of reference we have set out in Table~I the exponent
values adopted in the analyses reported below.  They correspond, of
course, to those for the three-dimensional Ising universality
class~\cite{Zinn98, Butera2000, Pelissetto2002, Guida1998,
Campostrini2002}.

Since extensive experimental data will be discussed it is useful to
specify here the units employed.  Thus: (i) mass densities will be
measured in $\gcm$ and denoted \stress{in} \stress{these} \stress{units}
as $\rrho \equiv \rho / (\gcm)$; (ii) for surface tensions,
$\Sigma$, and the corresponding amplitudes, $K$, etc., units of $\stu$
will be employed and indicated by $\rK$, etc.; (iii) since the reduced
free energy density will be taken in $\hbox{cm}^{-3}$ units, appropriate
units for the field $h$, conjugate to the order parameter $M = (\rho -
\rho_c)$, are $\hbox{g}^{-1}$ indicated correspondingly by $\rh$.  The
units for various critical amplitudes, $B$, $C^\pm$, and the
coefficients $\Sigma_{th}$ in \whatis{eq5.1:S0}, etc. follow similarly.

\begin{center}
{\bf Outline}
\end{center}

\noindent
The balance of this article is set out as follows.  In Sec.~II,
the theoretical predictions
\whatis{eq5.1:PQ} for the universal ratios $P$ and $Q$ are
discussed in the light of the experiments of
Woermann and
coworkers~\cite{aKreuser, Mainzer}.
The density data of Greer~\cite{Greer} along the coexistence curve
of isobutyric acid $+$ water
and the corresponding critical surface tension data of {\bf HWK}~\cite{Knobler}
are analyzed in Sec.~III to determine the critical amplitudes $B$ and
$K$.  These results are used
to calibrate the {\bf NWW} data which are much more extensive but not as precise and,
thus, harder to analyze with confidence on their own.
However, the {\bf NWW} observations are found to be fully consistent
with the earlier measurements and, furthermore, confirm the validity of
Antonow's law in the temperature range studied: see Fig.~2 below.

On this basis the analysis can be carried forward to determine the
crucial background term, $\Sigma_0(t,h)$.  The results along the
coexistence curve and on the critical isochore above $T_c$ (\ie, for
zero field, $h \negthinspace = \negthinspace 0$) are reported in
Sec.~IV.A and displayed in Fig.~3.  To estimate $\Sigma_0(t,h)$ on the
critical isotherm, $T = T_c$ $(t=0)$ for nonzero $h$, it is necessary to
re-express the {\bf NWW} observations at constant density in terms of
the field.  For this purpose one needs information regarding the bulk
equation of state that goes beyond the value of $B$.  In the absence of
direct bulk measurement on the critical isotherm, one can progress by
appealing to bulk {\it universal amplitude ratios\/} (known numerically
with appreciable reliability~\cite{Zinn98, Butera2000, Pelissetto2002}),
combined with the further universal critical ratio $S^+$ that relates
the surface tension amplitude $K$ to the correlation length
\cite{Zinn98}: see (A.1) in the Appendix where the relevant theory is
reviewed and its application explained.

By this route one can estimate the amplitude $C^+$ for the ordering
susceptibility and thence, as explained in Secs.~IV.B, IV.C, and the
Appendix, the necessary conversions can be implemented.  Thereby the
field-dependence of the background may be gauged.  The singular part,
$\Delta \Sigma(t,h)$, then follows by subtraction although,
unfortunately, with rather limited precision that precludes, for
example, any direct determination of the surface tension amplitudes,
$K^c_>$ and $K^c_<$, \stress{on} the critical isotherms.  Nevertheless,
in Sec.~V, the theoretically predicted universal scaling functions
are compared in suitable plots with the
experimental surface tension data of {\bf NWW}.
To the extent that consistency is well established,
the results are encouraging: however, significantly
more precise and extensive data close to criticality will be required to
enable sharper tests of the theory.

The analysis, furthermore, encounters anew a puzzling dilemma concerning
the degree to which the dimensionless surface-tension ratio, $S^+$, for
the isobutyric acid $+$ water system actually conforms to the expected
value in light of independent observations of the critical scattering
and the values found for other systems.  The issue is brought up in
Sec.~IV.B and pursued in some detail in the Appendix although without
resolution.  The considerations presented highlight the need for further
experiments and, perhaps, for further theoretical developments.

The article is summarized briefly in Sec.~VI.

%
%
\section{Lutidine and Water: Universal Ratios $\boldsymbol P$ and
$\boldsymbol Q$}
\noindent
Predictions for the surface tension ratios $P$ and $Q$
defined in \whatis{eq5.1:PQ} have
been tested experimentally by Mainzer-Althof and Woermann
({\bf MW})~\cite{Mainzer}.
A binary liquid mixture of 2,6-dimethyl pyridine
[2,6-(CH$_3$)$_2$(C$_5$H$_3$N)], which is also known as 2,6-lutidine, and water
was
prepared at the critical composition and used in the experiment.
The surface tensions $\Sab(T)$, $\Sar(T)$, and $\Sabr(T)$
were measured, for the reduced temperature range $-0.028 \lsim t \lsim 0.019$, by
the inverted pendant drop (or rising bubble) method: analysis of
the contour line of the bubbles gives the surface tension.  Also, the mass
densities of the
liquid phases $\rho_\beta(T)$, $\rho_\gamma(T)$, and $\rho_{\beta\gamma}(T)$
were measured.  For the critical surface tension, {\bf MW} re-analyzed the
experimental data of Kreuser and Woermann~\cite{aKreuser} using the new
{\bf MW} density data.

In the {\bf MW} analysis of the surface tension data, the background
\whatis{eq5.1:S0} was employed truncated at the linear term, $\Sigma_t
\thinspace t$.  At first {\bf MW} fitted the data to \whatis{eq5.1:sabr} and
\whatis{eq5.1:sar} by
allowing separate background terms for $\Sab$, $\Sar$, and $\Sabr$ (\ie, different
sets of $\Sigma_c$ and $\Sigma_t$) and varying
$\Sigma_c$, $\Sigma_t$, $K^\pm$, and $(K+K^-)$.
For the exponent $\mu$ they assumed the value $1.26$.
The results did not agree
with the original Fisher-Upton ({\bf FU}) values
$P \simeq 0.1_2$ and $Q \simeq -0.83$ \cite{Fisher90a, Fisher90b}.
Furthermore, it appeared that Antonow's rule was \stress{not}
obeyed.  However, when the analysis was performed with, as demanded
theoretically,
a \stress{common} background, $\Sigma_0(T)$,
Antonow's rule was found to be satisfied (as expected): this procedure yielded
\begin{equation}
\Pexpt \simeq 0.00 \pm 0.08, \quad
\Qexpt \simeq -1.0 \pm 0.2.
\label{eq5.2:fit1}
\end{equation}
The $P$ ratio is evidently not quite consistent with the {\bf FU} value
but, significantly, the $Q$ value is consistent, both in sign \stress{and}
magnitude.

Other fitting procedures were also explored by {\bf MW} by imposing the
theoretical value of \stress{either} $K^+/K \enspace [= PQ/(1+Q)]$ \stress{or}
$K^-/K \enspace [= P/(1+Q)]$ and using a common background.  Averaging the
central values of two different {\bf MW} fits and extending error estimates
to fully cover the fits yield
\begin{equation}
\Pexpt \simeq 0.11^{+0.05}_{-0.08}, \quad
\Qexpt \simeq {-0.835}^{+0.06}_{-0.12}.
\label{eq5.2:fit2}
\end{equation}
These values are consistent both with \whatis{eq5.2:fit1} and, now, with the
original {\bf FU} estimates.  Furthermore, the more recent improved EdGF estimates
for $P$ and $Q$ from {\bf I}, quoted above in 
\whatis{eq5.1:PQ}, agree remarkably well with these albeit biased fits to the
experimental data.

It is appropriate to mention some earlier work.
Pegg, Goh, Scott, and Knobler~\cite{Pegg85} studied quasi-binary mixtures of
$n$-octadecane and $n$-nonadecane in ethane and measured the surface tensions
through and near both the \stress{upper} and the \stress{lower}
critical endpoints that occur in
the vicinity of the \stress{hidden} \stress{tricritical}
\stress{point} that arises in this system.
Their data, although limited, certainly support Antonow's rule and, more
importantly for us, suggest strongly that the ratio $K^+/K$ is negative and of
order unity in accordance with the EdGF predictions.  However, the
upper and lower endpoints are less than $0.2 \thinspace \celc$ apart and,
mainly, for that reason, the 8 data points for the surface tensions $\Sbr$
(and the 3 for $\Sar$) are not sufficient to warrant quantitative study.

The surface tension of the water and 2,5-lutidine system has been measured by
Privat and co-workers~\cite{Privat} at the critical composition of the lower
critical consolute point as well as off that critical composition.  The authors
claim to verify Antonow's rule.  However, their quoted fits for the amplitudes
$K$, $K^+$, and $K^-$ seem problematical relative to the graphical
presentation of their data.  Thus, and especially in the light of the
{\bf MW}
experiments on the closely related 2,6-lutidine system, this work cannot be
regarded as yielding useful experimental estimates for the amplitude
ratios~$P$ and~$Q$.

%
%
\section{Isobutyric Acid and Water: Nonuniversal Amplitudes}
\subsection{Coexistence Curve Amplitude}
\noindent
The coexistence curve of isobutyric acid $+$ water has been studied
carefully by Greer~\cite{Greer}, who measured the coexisting mass
densities, $\rho_\beta(T)$ and $\rho_\gamma(T)$, with a precision
of 20 ppm from 3.5 K below the critical temperature at intervals
as small as 5 mK (with a precision of $\pm 1$ mK) for a sample
prepared at close to the critical composition.  Two runs were made
and the data are recorded in Appendix A of Ref.~\onlinecite{Greer}.
Greer's analysis demonstrated that an optimal choice of order parameter
(as judged in relation to the number density of a pure fluid in a
lattice-gas representation) was the volume fraction of one component,
say $\phi_B$.  The surface tension experiments of {\bf NWW}, which are
of principal concern to us, however, also used the mass density in the
vicinity of the critical endpoint as the primary controlled observable.
Accordingly, we have reanalyzed Greer's data using the mass difference,
$\rho - \rho_c$, as the order parameter.  This choice, of course,
affects the real (and apparent) magnitudes of the various correction
terms; but these will play no more than an auxiliary role in our
analysis.  Further discussion of fits for the volume fraction
difference, $\Delta \phi$, are given in \cite{Zinn1997}: one learns that
a range of reasonable fits to the data serve to determine the critical
point to within $\pm 10$ to $15$ mK while the leading critical amplitude
$B$ can be found with a precision of no better than $\pm 1$ \%.

More specifically, we have fitted the data to
\begin{equation}
\hbox{$1\over2$} \Delta \rho
\equiv \hbox{$1\over2$} \left [ \rho_\gamma(T) - \rho_\beta(T) \right ]
= B |t|^\beta \thinspace
[ 1 + b_\theta |t|^\theta + b_1 t], \label{eq5.3:form}
\end{equation}
with $t = (T-T_c)/T_c$ as usual, but, as a useful crosscheck, also
in terms of the asymptotically equivalent variable,
\begin{equation}
t' = 1 - T_c/T = t / (1+t), \label{eq5.3:tp}
\end{equation}
as in a previous study \cite{Zinn}.  Using primes to denote the
amplitudes fitted with $t'$ we find, from the first run \cite{Zinn1997},
\begin{eqnarray}
&&T_c \simeq 25.996_7\celc, \quad \check B \simeq \check B' \simeq
0.03104, \nonumber\\
&&b_\theta \simeq b'_\theta \simeq -1.73, \quad
 b_1 \simeq -9.783, \quad b'_1 \simeq -9.455.\label{eq5.3:Brho}
\end{eqnarray}
The difference $b'_1 - b_1 \simeq 0.328$ is close to $\beta$ as it
should be for consistency in light of \whatis{eq5.3:tp}.

The exponent $\theta \simeq 0.5$ (See Table~I) in \whatis{eq5.3:form}
specifies the leading correction to scaling.  General scaling
considerations \cite{Fisher2000, Kim1998} indicate that a further
correction term, $|t|^{2\beta}$, should also be present which, since
$2 \beta \simeq 0.65 < 1$, should in fact dominate the linear term.  The
corresponding amplitude $b_{2\beta}$ may well be rather small:
nevertheless, the data cannot resolve two such terms so that the fitted
values of the coefficients $b_\theta$ and $b'_\theta$ must be regarded
as no more than \textit{effective} \textit{amplitudes}.

On setting $b_\theta = b_1 = 0$ in \whatis{eq5.3:form} Greer found
$\check B = 0.0265 \pm 0.0005$; however, on allowing for $b_\theta \ne
0$ the value $0.0315 \pm 0.0025$ resulted; this is fully consistent with
\whatis{eq5.3:Brho}.
Repeating the same procedure for the second experimental run yields
\begin{eqnarray}
&&T_c \simeq 25.969_6 \celc, \quad \check B \simeq \check B' \simeq
0.03175, \nonumber\\
&&b_\theta \simeq b'_\theta \simeq -2.635, \quad
  b_1 \simeq -14.563, \quad b'_1 \simeq -14.234.\label{eq5.3:Brho2}
\end{eqnarray}
The difference $b'_1 - b_1 \simeq 0.329$ is again close to $\beta$.
Clearly, the deviations of the fitted values here from those in
\whatis{eq5.3:Brho} provide a measure of the accuracy available in
estimating $T_c$ and $B$.

%
%
\subsection{Comparison with NWW Coexistence Curve Data}
\noindent
We may now use the fits to \whatis{eq5.3:form} to calibrate the
experiments
on the same system by {\bf NWW} who prepared mixtures of
isobutyric acid $+$ water at various compositions and determined
the surface tensions
by optically measuring the wavelength of surface waves generated
by a transducer.  In Table~1 of {\bf NWW}, about 140 $(T, \rho, \Sigma)$
data
points are presented: these are what we study here.  We have
extrapolated some of the observations to the coexistence curve;
but, as indicated briefly in Sec.~IV.A, only very small
changes in the values of $\rho$ and $\Sigma$ are entailed.

To proceed, we first consider the diameter of the coexistence curve in the
density variable, namely,
\begin{equation}
{\overline \rho}(T) = \hbox{$1\over2$} (\rho_\beta + \rho_\gamma),
\label{eq5.3:diameter}
\end{equation}
since, as mentioned, the density was directly observed by {\bf NWW}.
Owing to the mixing of thermodynamic fields near criticality in fluids,
one expects a dominant singular term $|t|^{1-\alpha}$ to appear in the
diameter~\cite{Fisher2000, Kim1998, Widom70, Rehr71, Fisher75}.
However, the {\bf NWW} data do not reveal a signature of any such singular term.
Rather, the coexistence curve appears almost symmetric in the
$(\rho, T)$ plane \cite{GreerDiameter}.  Thus the diameter
may be well represented by a constant as
\begin{equation}
{\check {\overline \rho}}(T) \simeq \check \rho_c \simeq 0.9936.
\label{eq5.3:rhoc}
\end{equation}

\begin{figure}
\includegraphics{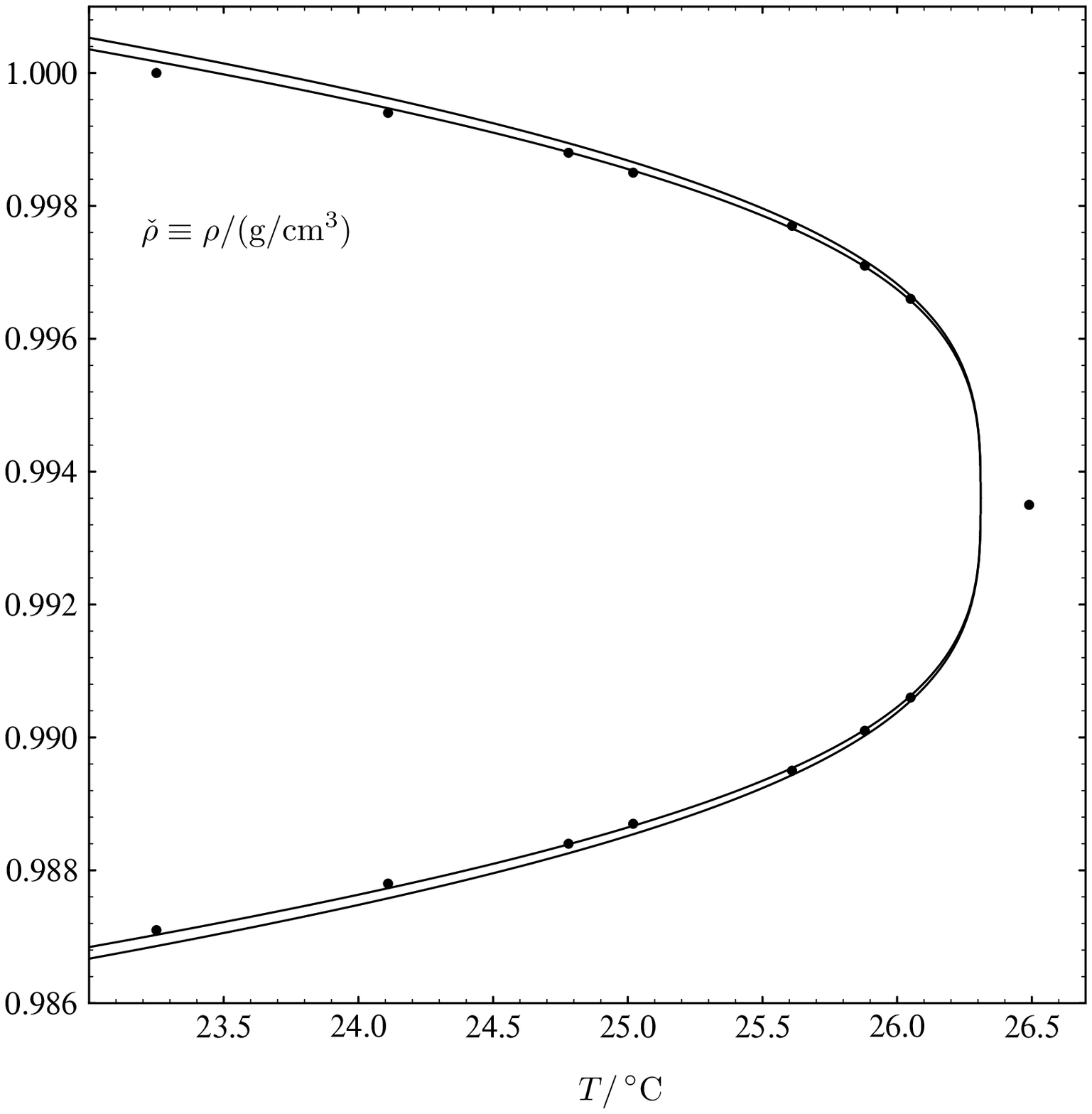}
\caption{\label{fig5:coex}
}
\end{figure}

As discussed by Greer \cite{Greer}, the critical temperature of the
isobutyric acid $+$ water system is particularly sensitive to ionic
impurities \cite{Kim2001} and, indeed, 
experience shows that the observed critical temperatures of
nominally the same binary
fluid mixtures typically vary from experiment to
experiment by amounts exceeding the stated errors.  For
fitting the {\bf NWW} data, therefore, we have adopted
their value of $T_c = 26.310 \pm 0.001 \celc$,
even though it 
lies outside the uncertainty limits implied by \whatis{eq5.3:Brho}
and \whatis{eq5.3:Brho2}.
As seen in Fig.~\ref{fig5:coex}, by using \whatis{eq5.3:form} and
\whatis{eq5.3:rhoc} with the amplitude estimate
\begin{equation}
\check B = 0.0314 \pm 0.0004,\label{eq5.3:amplB}
\end{equation}
the {\bf NWW} data can be represented
rather satisfactorily.  This demonstrates the consistency of their
observations with other careful studies and establishes a reliable
estimate for the amplitude $B$ which will be employed below.

It may be remarked that {\bf NWW} advocated the use in theoretical analysis
of a
(conventionally defined) volume fraction, $\phi$, in place of
the density, $\rho$ \cite{Phi}.  Study of their data reveals, however, that the
dependence of $\phi$ on $\rho$ is both surprisingly nonlinear and
significantly temperature
dependent.  Furthermore, contrary to the expectations of {\bf NWW}
(see p.~5781) and Greer, the coexistence curve appears
to be more symmetric and
regularly behaved in terms of $\rho$ than of $\phi$.  Accordingly, we have
accepted the directly observed variable $\rho$ 
as the order parameter in all the following analysis.

%
%
\subsection{Critical Surface-Tension Amplitude $\boldsymbol K$}
\noindent
In order to obtain the critical amplitude $K$ as defined for the surface
tension in \whatis{eq5.1:sbr},
the capillary-rise data obtained by {\bf HWK} \cite{Knobler}
for isobutyric acid $+$ water will be employed.
In their Fig.~8, one can clearly
see one point that deviates significantly from the general trends, namely,
that corresponding in {\bf HWK} Table~VII to $T_c - T = 1.088$ K or
$t \simeq -0.0036$, and $\check \Sigma_{\beta|\gamma} = 0.0191$.
This point has thus been omitted in the present analysis.

Since only eight data points are available, the correction-to-scaling terms
are not expected to play a detectable role.  First, therefore,
the data have been fitted to the form
\begin{equation}
\Sigma_{\beta|\gamma} = \Sigma_c + K |t|^\mu,
\label{5.3:sbr}
\end{equation}
where, of course, $\Sigma_c \equiv 0$ is anticipated.
The reported value of
$T_c$ is $299\thinspace\kelvin$ \cite{Knobler}; but we have not found
in the account of {\bf HWK} a satisfying description as to how this
value was determined.  We therefore adopted
\begin{equation}
T_c = 299.00 \thinspace \kelvin = 25.85 \thinspace \celc, 
\label{eq5.3:KnoblerTc}
\end{equation}
provisionally but have also examined how estimates for $K$ are affected by
changes in the assigned value of $T_c$.

A weighted least-squares fit then yields
\begin{equation}
\check \Sigma_c = -0.002 \pm 0.008, \quad \check K = 34.6 \pm 5.7,
\label{eq5.3:roughK}
\end{equation}
where the uncertainties quoted here and below are twice the standard deviation.
By imposing $\Sigma_c = 0$, which is clearly consistent with the data, one finds
\begin{equation}
\check K = 32.8 \pm 1.7, \label{eq5.3:sc0}
\end{equation}
where the central estimate has shifted by 5\% while the uncertainty is reduced
by 30\% in comparison with \whatis{eq5.3:roughK}.  Repeating the analysis
with $\Sigma_c = 0$ but using $t' = t/(1+t)$ as a variable \cite{Zinn}, we find
$\check K = 32.6 \pm 1.7$ in good agreement.

As a further check one can calculate $\Sigma_{\beta|\gamma} /
|t|^\mu$ from each data point.  This ratio approaches $K$ when $t\to 0$,
and an average yields $\check K \simeq 32.7$, suggesting that the corrections
to the pure power law are relatively small.

To gauge the sensitivity to the imposed value of $T_c$, we changed the value
by $\pm 0.1$ K.  Considering that the temperature was controlled to within
$0.01$ K \cite{Knobler}, this variation is fairly large.  The results
(obtained by fixing $\Sigma_c = 0$) are $\check K = 30.6 \pm 1.7$ for
$T_c = 25.95 \thinspace \celc$ and $\check K = 35.3 \pm 1.8$ for
$T_c = 25.75 \thinspace \celc$.
Despite the changes of $\pm 7$ \%, the overlap of the 
error bars allows consistency with the previous
estimates for $K$.

Finally, the data were fitted to $\Sigma = K \thinspace |(T/T_c) - 1|^\mu$
treating both $T_c$ and $K$ as parameters.  Minimizing $\chi^2$ yielded
$\check K \simeq 33.8_5$ and $T_c \simeq 25.78 \thinspace \celc$.
This value for $T_c$ is $0.07$ K lower than the asserted value
\whatis{eq5.3:KnoblerTc} but 
the amplitude estimate again lies in the indicated range.
Repeating the procedure with $\Sigma = K' \thinspace |1 - (T_c/T)|^\mu$,
yields the same value for $T_c$ but $\check K' \simeq 33.5$,
about 1\% smaller.  Overall we believe that
\begin{equation}
\check K = 33.7 \pm 2.5, \label{eq5.3:Kexp}
\end{equation}
summarizes the evidence conservatively \cite{Kprob}, although a more precise
knowledge of $T_c$ in the {\bf HWK} experiments could be valuable.

%
%
\subsection{Comparison with NWW Critical Surface-Tension Data}
\noindent
It is informative to compare the experiments performed by {\bf HWK}
with those of {\bf NWW} to check, in particular, for
mutual consistency: see Fig.~\ref{fig5:chk}.  For convenience
{\bf NWW} data points on
the coexistence curve have been numbered consecutively from~1 to~15.
Except for point~8, other points joined by parallelograms were measured at the
 same temperature.  Points~1--7 lie on the high density side of the
coexistence curve and hence represent $\Sar(T,
h\negthinspace=\negthinspace0+)$, while points~9--15 correspondingly lie on
the low density side and represent $\Sab(T, h\negthinspace=\negthinspace0-)$.

To explain the significance of the fixed-$T$ parallelograms,
consider, as an example,
points~2 and~14.  By Antonow's rule we have $\Sar = \Sab + \Sbr$
and so by using
$\Sbr(T) \approx K|t|^\mu$ with the estimate \whatis{eq5.3:Kexp}
and by varying $\Sab$ within the uncertainty limits, one can find
a range of $\Sar$ values that are consistent with Antonow's rule.
Evidently, 
the central value of the data point~14 corresponds to the lower limit of the
data point~2, while the upper limit corresponds to the central value of data
point~2.  These mutually consistent limiting pairs of
the data points~2 and~14 have been joined by parallel lines.
In other words, the vertical sides of the
parallelogram connecting data points~2 and~14 represent error limits on
the {\bf NWW} data that are consistent with
the {\bf HWK} experiment and with Antonow's rule.  Thus we notice that
data point~3 represents a high estimate of $\Sab$ while 
data point~13 is relatively low.  Despite this (3,13) worst case,
it can be concluded that the two experiments are mutually consistent and that
the estimate~$K$ in~\whatis{eq5.3:Kexp} is reliable \cite{Kprob} .
Note that it would have
been very difficult to obtain a reliable value of $K$ from the {\bf NWW} data alone,
owing to the necessity of extrapolating and differencing their (not highly
precise) data to obtain $\Sbr$ via Antonow's rule.

%
%
\subsection{Other Surface Tension Amplitudes}
\noindent
Having determined the two amplitudes $K$ and $B$, which serve as
metrical factors in the scaling formulation of the critical-endpoint
behavior of the surface tension [see \whatis{eq5.1:Spm}], we are
prepared to take the next step towards analyzing the {\bf NWW} data.
The first issue must be to establish the background contribution,
$\Sigma_0(T,h)$, by estimating the coefficients $\Sigma_c$, $\Sigma_t$,
etc., in \whatis{eq5.1:S0}.  To that end it proves helpful to have
at hand values for the surface tension amplitudes $K^+$ and $K^-$ as
defined in \whatis{eq5.1:sabr}.  Ideally, of course, these should be
unambiguously determined by the data themselves: but, as already seen in
Sec.~II, it is unrealistic at this stage in the development of
experimental techniques to expect to do more than
verify---optimistically at a fully convincing
level---\stress{consistency} of the theory with the observational data.
Accordingly, we report here the values 
\begin{equation}
\check K^+ = -23.3 \pm 1.8, \quad \check K^- = 27.9 \pm 2.1,
\label{eq5.3:Kpm}
\end{equation}
that follow from \whatis{eq5.3:Kexp} on accepting the EdGF calculations for
the universal ratios $P$ and $Q$ given in \whatis{eq5.1:PQ}.

To determine the field-dependent coefficients $\Sigma_h$ and
$\Sigma_{hh}$ in the background expansion \whatis{eq5.1:S0} we will need
to examine the data for $\Sigma(t,h)$ on the critical isotherm.  The
expected behavior is
\begin{equation}
\Sigma_c(h) \approx K^c_{\gl} |h|^{\mu/\Delta} + \Sigma_c + \Sigma_h h +
	\Sigma_{hh} h^2, \qquad (T = T_c)
\label{eq5.3:TcS0}
\end{equation}
where, again, values for the amplitudes $\check K^c_>$ and $\check
K^c_<$, for $h \gl 0$, respectively, will prove of importance.  To
estimate these one may appeal to the EdGF theoretical values reported in
{\bf I} (6.9) for the universal ratios
\begin{equation}
S^c_{\gl} \equiv {K^c_{\gl} \over K} \left ( {B \over C^+} \right )
^{\mu / \Delta}.
\label{eq5.3:Scgl}
\end{equation}
In the second factor in this expression, which enters because $K^c_{\gl}$
and $K$ have different dimensions, the nonuniversal amplitude $C^+$
specifies the magnitude above $T_c$ of the divergence of the basic
ordering susceptibility as $1/t^\gamma$.  Note that the notation for
amplitudes used here and below accords with that set out in the Appendix
of \cite{Zinn98}.

The appropriate value of $C^+$, and of other nonuniversal amplitudes for
isobutyric acid $+$ water, is taken up in Sec.~IV.B when the
field-dependence is considered explicitly.  At that stage, we will make
contact with Moldover's extensive analysis \cite{Moldover85, Moldover86}
of surface tension, light scattering, and bulk thermodynamic experiments
for a range of single-component and binary fluid systems designed to
test the hyperscaling (or ``two-scale factor universality'') hypothesis
for criticality.  Various details and some questions they raise are
presented in the Appendix.

%
%
\section{Determination of the Surface Tension Background}
\noindent
As discussed in Sec.~I, to analyze the {\bf NWW} surface
tension measurements appropriately,
the background~$\Sigma_0(T,h)$ must at least contain quadratic terms in $t$;
it then seems reasonable to include quadratic terms in $h$ 
also as in \whatis{eq5.1:S0}.
In order to determine the background
coefficients $\Sigma_c$, $\Sigma_t$, etc., 
we adopt the following strategy.  First, the data on the
$h\negthinspace=\negthinspace0$
axis, \ie, for the mixture at the critical composition,
will be fitted to obtain
$\Sigma_c$, $\Sigma_t$, and $\Sigma_{tt}$.
From the data on the critical isotherm, $t=0$,
the coefficients $\Sigma_h$ and $\Sigma_{hh}$ can then be examined.
Finally, to obtain the cross-term
coefficient $\Sigma_{th}$, the derivative
$(\partial \Sigma / \partial t)_h$ may be studied
on the critical isotherm.

%
%
\subsection{Along the Coexistence Curve}
\noindent
In zero-field, the surface tension above $T_c$ is,
following the Introduction, expected to behave as
\begin{equation}
\Sabr(T) \approx K^+ |t|^\mu + \Sigma_c + \Sigma_t t + \Sigma_{tt} t^2,
\qquad (T \ge T_c).
\label{eq5.4:sabr}
\end{equation}
Below $T_c$, there are two distinct vapor-liquid interfaces
and two surface tensions, $\Sab$ and $\Sar$, corresponding to the
coexisting phases $\beta$ and $\gamma$.
By Antonow's rule (which as shown is
satisfied by the {\bf NWW} data) the difference $(\Sar - \Sab)$
is the critical surface tension $\Sbr$ and thus it suffices to study
only $\Sab(T)$ below $T_c$.
The identical expression \whatis{eq5.4:sabr} then applies except that
$K^+$ is to be replaced by $K^-$.

To proceed, one may first fit $\Sabr(T)$ using
the data above $T_c$ at the very close-to-critical composition
tabulated by {\bf NWW}, by regarding
$K^+$, $\Sigma_c$, $\Sigma_t$, and $\Sigma_{tt}$ as free parameters.
These fits suggest $K^+ > 0$ which, in accord with the generally observed
upwards curvature (compare with {\bf I} Fig.~5),
has the \textit{opposite} sign to the prediction for $K^+$ in
\whatis{eq5.3:Kpm}.
We conclude that the {\bf NWW} data above $T_c$ are not, on their own,
of adequate
precision to reliably distinguish the amplitudes of three terms varying as $t$,
$t^\mu$, and $t^2$; hence, we cannot independently estimate $K^+$.

However,
as indicated in the {\bf MW} analysis \cite{Mainzer} discussed in
Sec.~II, the situation may
be improved by imposing the theoretical value of the ratio $K^+/K$;
but this is then
equivalent to imposing $K^+$ because, as seen in Sec.~III,
the critical amplitude $K$ has been already estimated reliably
by comparison with
the {\bf HWK} data.  Since $\Sabr(T)$ and $\Sab(T)$ share the
common background $\Sigma_0(T,0)$, we likewise impose $K^-$ as given
in \whatis{eq5.3:Kpm}.
Of course, by adopting this procedure we will be primarily checking the
\stress{consistency} of our theory with the {\bf NWW} data which,
unfortunately, are not
adequate for estimating $K^+$ and $K^-$ directly as might originally
have been hoped.

Accordingly,
by subtracting the singular term in \whatis{eq5.4:sabr} and likewise
below $T_c$,
we can generate experimental data for the zero-field background
alone: see Fig.~\ref{fig5:sigma0}.
Again we use the {\bf NWW} data taken at the (close-to) critical composition above
$T_c$ and the (close-to) low-density side of the coexistence curve.
A least-squares fit then yields
\begin{equation}
\check \Sigma_c \simeq 26.284, \quad
\check \Sigma_t \simeq -14.6, \quad
\check \Sigma_{tt} \simeq 381.
\label{eq5.4:h0fit}
\end{equation}
The corresponding plot is represented by the dashed curve
in Fig.~\ref{fig5:sigma0} and
seen to be very good.  This demonstrates
that these coefficient estimates are most reasonable although, as will be
discussed, they prove to be less than fully satisfactory in the near
critical region below~$T_c$.

To examine the fit further
consider the plots shown in Fig.~\ref{fig5:chk}.
The dashed curve represents the surface tensions $\Sab(T)$ and $\Sar(T)$
on the coexistence
curve as reproduced from the amplitudes $K$ and $K^-$ and the background
coefficients \whatis{eq5.4:h0fit}: to translate $t \propto T_c - T$ into
the density on the coexistence curve, the inner fit described in the
caption of Fig.~\ref{fig5:coex} is used.  The agreement with the central
values of the {\bf NWW} data on the low-density side
is encouraging;
however, the fit reveals significant, systematic discrepancies on the
high density side where the plot lies outside the observational range for all
the points 1--7.

To complement the overall or ``full-range''
approach just described and to understand the
discrepancies uncovered on the high-density side of the coexistence curve,
let us
return to Fig.~\ref{fig5:chk} and focus on the {\it data closest to
criticality\/}.
In particular, as explained in the construction of the parallelograms in this
figure, consistency with the {\bf HWK}
data for $\Sbr(T)$, together with Antonow's
rule~\cite{Fenzl} requires that the allowed observational uncertainties
quoted for the {\bf NWW}
data should be reduced to correspond to the vertical sides
of each parallelogram.  Accordingly, for $T \lsim T_c$
consider the truncated forms
\begin{eqnarray}
\Sab(T) &\simeq& K^-|t|^\mu + \Sigma_c + \Sigma_t t, \label{eq5.4:linSab}\\
\Sar(T) &\simeq& (K^-+K) |t|^\mu + \Sigma_c + \Sigma_t t,\label{eq5.4:linS0}
\end{eqnarray}
in which the background has been restricted to a linear term.
By trial and error, one can discover those values of $\Sigma_c$ and
$\Sigma_t$ that give acceptable fits to those data points lying closest
to the critical point, specifically, for the pairs
(7,9), (6,10), (5,11), (4,12), and
(3,13).  Slight shifts in the nominal value of $\rho_c$ from
\whatis{eq5.3:rhoc} may also be examined but are
found to be detrimental for changes
exceeding $\Delta \check \rho_c = 10^{-4}$.

The predominant conclusion is that $\Sigma_c$, the common value
of the surface tensions $\Sab$, $\Sar$, and $\Sabr$ \stress{at} criticality,
cannot actually be as low as suggested by the full-range
fit \whatis{eq5.4:h0fit}.
The acceptable ``critical fits'' yield, instead, the range
\begin{equation}
\check \Sigma_c = 26.294 \pm 0.004.
\label{eq5.4:Sc}
\end{equation}
Indeed, {\bf NWW} actually quote the estimate
$\check \Sigma_c \simeq 26.30$, which is
much closer to this ``critical value'' than to \whatis{eq5.4:h0fit}.  The
corresponding
preferred value for the background slope, is found to be
$\check \Sigma_t = -17.5$; but reasonable fits could be obtained in the range
$-16.5 > \check \Sigma_t > -18.5$.  In all these fits,
however, the fitting curve in Fig.~\ref{fig5:chk} became too broad at the larger values of $| \rho -
\rho_c |$, failing to fit the data-point pairs (1,15) and (2,14).  This last
defect, however, was overcome by including in 
\whatis{eq5.4:linSab} and \whatis{eq5.4:linS0} the
quadratic term $\Sigma_{tt} t^2$.  As the assigned value $\Sigma_{tt}$ is
increased, keeping $\Sigma_c$ fixed as proves essential, the optimal value for
$\Sigma_t$ correspondingly falls.  Thus $\check \Sigma_t = -16.0$ and
$\check \Sigma_{tt} = 250$ give very good fits in
Fig.~\ref{fig5:chk}.  However, these particular
values do not describe well the
highest data points \textit{above} $T_c$ in Fig.~\ref{fig5:sigma0} (for $t >
0.015$).  One could well disregard this range since the data extend
below $T_c$ only down to $|t| \simeq 0.011$.  However, one can find
values that prove rather satisfactory in fitting \stress{all\/} the $h=0$
data, \ie, on the coexistence curve and at the critical composition
above $T_c$, namely,
\begin{equation}
\check \Sigma_c \simeq 26.294, \quad
\check \Sigma_t \simeq -15.0, \quad
\check \Sigma_{tt} \simeq 400.
\label{eq5.4:h0fitb}
\end{equation}
These coefficients yield the solid curve shown in Fig.~\ref{fig5:sigma0}:
evidently the fit displays a systematic displacement
above the central values of the
data points but by rather small amounts that remain \stress{within}
all the uncertainty ranges.
Furthermore, the agreement with the coexistence data
is now much improved as seen by the solid curve in Fig.~\ref{fig5:chk}.
Only very minor departures from the acceptable uncertainty ranges
now arise.  Accordingly we will
retain the assignment \whatis{eq5.4:h0fitb} in all the following analysis.

At this stage we may conclude that the {\bf NWW}
data are, first, fully consistent
with the best previous observations of the coexistence curve
and the critical surface tension and, second,  
are quite consistent with the EdGF based estimates
of the ratios $K^+/K$ and $K^-/K$ even though,
unfortunately, the data cannot critically test these ratios.

%
%
\subsection{On the Critical Isotherm}
\noindent
It transpires, in retrospect, that {\bf NWW}
did not make observations of the surface tensions
very close to the critical
isotherm $T=T_c = 26.310 \thinspace \celc$.  We may, however, presume
that the isotherms just
above and below $T_c$, namely, at $T = 26.49 \thinspace \celc$ and
$T = 26.05 \thinspace \celc$ constitute rough lower and upper bounds,
respectively, on $\Sigma_c(h)$.
On the critical isotherm, the surface tension
should behave as in \whatis{eq5.3:TcS0} with $\mu / \Delta \simeq 0.806$.
The optimal assignment of the critical point value $\Sigma_c$ has already
been determined in the previous subsection.  One might wish to determine
the amplitudes $K^c_>$ and $K^c_<$ directly from the data; but,
regrettably there is, once again, no hope of that!  Rather, with the
limited aim of testing only consistency with the theory, we will adopt
theoretical estimates for these two amplitudes.  At this point one
encounters a rather sharp and surprising \stress{dilemma}: as explained
further in the Appendix, two significantly different sets of EdGF-based
predictions arise from \whatis{eq5.3:Scgl} and {\bf I} (6.9).  The first
set, derives \stress{entirely} from the two metrical factors $K$ and
$B$, that we have determined for isobutyric acid $+$ water, combined
with well-confirmed values for various {\it universal critical amplitude
ratios\/} \cite{Zinn98}: as explained in the Appendix this yields
\begin{equation}
\rlap{\hskip -.1\displaywidth\hbox{{\bf A}:}}
\check K^c_> = (5.0 \pm 0.6) \negthinspace \times \negthinspace 10^{-18},
\quad\quad\quad
\check K^c_< = -(1.5_5 \pm 0.1_7) \negthinspace \times \negthinspace 10^{-18}.
\label{eq5.4:KcglA}
\end{equation}

The second set also utilizes the estimate \whatis{eq5.3:Kexp} for $K$
in \whatis{eq5.3:Scgl} together with \whatis{eq5.3:amplB} for $B$; but
instead of using the purely theoretically derived value, namely,
\begin{equation}
\rlap{\hskip -.286\displaywidth\hbox{{\bf A}:}}
\check C^+ = (3.04 \pm 0.30) \negthinspace \times \negthinspace 10^{-26},
\label{eq5.4:CpA}
\end{equation}
for the amplitude $C^+$ in \whatis{eq5.3:Scgl}, it employs the much larger
estimate
\begin{equation}
\rlap{\hskip -.291\displaywidth\hbox{{\bf B}:}}
\check C^+ = (14.2 \pm 0.6) \negthinspace \times \negthinspace 10^{-26},
\label{eq5.4:CpB}
\end{equation}
that follows, as explained in the Appendix, from {\it independent\/}
light scattering and thermodynamic experimental data for isobutyric acid
$+$ water \cite{Moldover85, Moldover86}.  This then leads to the
alternative estimates
\begin{equation}
\rlap{\hskip -.103\displaywidth\hbox{{\bf B}:}}
\check K^c_> = (17.5 \pm 1.5) \negthinspace \times \negthinspace 10^{-18},
\quad\quad\quad
\check K^c_< = -(5.4 \pm 0.5) \negthinspace \times \negthinspace 10^{-18},
\label{eq5.4:KcglB}
\end{equation}
that, likewise, are much larger than in \whatis{eq5.4:KcglA}.

Granted the values of $K^c_{\gl}$ in the critical point background
\whatis{eq5.3:TcS0}, one is left with only
the coefficients $\Sigma_h$ and $\Sigma_{hh}$ to be determined
by fitting.  However,
since the {\bf NWW} measurements were made in terms of the density,
it is necessary
to convert the $(t, M \negthinspace =  \negthinspace\rho \negthinspace -
\negthinspace\rho_c)$ data to $(t, h)$ using an equation
of state.  To this end, it is natural to employ the 
scaled equation of state derived in \cite{Zinn99} on the basis of
the extended sine model.  The metric factor $B$ given in
\whatis{eq5.3:amplB} and the appropriate values of $C^+$ [in
\whatis{eq5.4:CpA} or \whatis{eq5.4:CpB}] may be used.
The data spread over the range
$-4.8 \times 10^{21} \lsim \check h \lsim 2.4 \times 10^{20}$
[using \whatis{eq5.4:CpA}] and, hence, reach more widely
on the negative $h$ (or low density)
side; but for fitting we will focus on the available symmetric range.

On the basis of the {\bf A} values above we thus
find that the assignment
\begin{equation}
\check \Sigma_h \simeq 3.72 \negthinspace \times \negthinspace 10^{-23}
\quad \quad \quad \hbox{and} \quad \quad \quad
\check \Sigma_{hh} \simeq 6.65 \negthinspace \times \negthinspace 10^{-43},
\label{eq5.4:Tcfits}
\end{equation}
provides a fit that lies well between the upper and lower limits
for $|h|$ not too large: see Fig.~\ref{fig5:attc}.
Comparable fits can be found using the {\bf B} values
but with, of course, substantially different values of $\Sigma_h$ and
$\Sigma_{hh}$ and a different accessible range of $h$.  For the balance of
this article, however, we will stay with the `purely theoretical' {\bf A}
estimates \whatis{eq5.4:KcglA} and \whatis{eq5.4:CpA}.

%
%
\subsection{Cross Term}
\noindent
In order to estimate the cross coefficient $\Sigma_{th}$ in the
background \whatis{eq5.1:S0}, one should examine
the derivative $(\partial \Sigma / \partial
t)_h$ at $T=T_c$.  Since the measurements were made
at constant $M$ (or density), it is
appropriate to invoke the relation
\begin{equation}
\left( \PD{\Sigma}{t} \right)_{\negthinspace\negthinspace h} =
	\left(\PD{\Sigma}{t}\right)_{\negthinspace\negthinspace M}
	+ \left( \PD{\Sigma}{M} \right)_{\negthinspace\negthinspace t}
	  \left( \PD{M}{h} \right)_{\negthinspace\negthinspace t}.
\label{eq5.4:xrel}
\end{equation}
However, owing to the measurement uncertainties, the estimates of
$(\partial \Sigma / \partial t)_M$ obtainable from the
{\bf NWW} data are extremely noisy.
Hence, one cannot seriously attempt to estimate $\Sigma_{th}$.
One may reasonably guess, nevertheless, that the geometric mean
of $\Sigma_{tt}$ and $\Sigma_{hh}$, namely,
using \whatis{eq5.4:h0fitb} and \whatis{eq5.4:Tcfits},
\begin{equation}
\left ( \check \Sigma_{tt} \check \Sigma_{hh} \right)^{1/2}
= 1.6 \negthinspace \times \negthinspace 10^{-20},
\label{eq5.4:S2x}
\end{equation}
will 
provide some indication of the possible order of magnitude of the
cross-coefficient.  However, 
we comment below on the sensitivity of the final
comparisons with experiment to the value of
$\Sigma_{th}$.

%
%
\section{Scaling Functions for the Surface Tension}
\noindent
Having found, to the somewhat limited degree feasible,
a reasonable description of the important surface-tension
background, we may examine the experimental data for consistency with the
predicted universal scaling functions $S_M^\pm(\mt)$
introduced in \whatis{eq5.1:Spm}.

In order to subtract the background $\Sigma_0(t,h)$, from the full surface
tension data, it is necessary, as before,
to convert the variables $(t, M \negthinspace = \negthinspace \rho
\negthinspace -  \negthinspace \rho_c)$ to $(t, h)$
because the experiments employed the density as a primary variable.
Adopting the {\bf NWW} critical
temperature $T_c = 26.310 \thinspace \celc$ and using $B$ from
\whatis{eq5.3:amplB}, the first step is to compute
$\mt = \penalty-100000 (\rho - \rho_c)/B|t|^\beta$.
The universal, scaled equation of state $\mt = Q_\pm(\tilde h)$ [for
$t \gl 0$] may be used in the form of the extended sine model of
\cite{Zinn99} to implement the conversion to $\tilde h$.
The corresponding $h$ follows via
$h = B \tilde h |t|^\Delta / C^+$ where, for the present discussion we
use \whatis{eq5.4:CpA} for $C^+$ as in Fig.~\ref{fig5:attc}.
Then, using the background with the coefficients given
in \whatis{eq5.4:h0fitb} and \whatis{eq5.4:Tcfits}, an estimate
for the singular part of the surface tension, $\Delta \Sigma (T, h)$,
can be extracted from the {\bf NWW} data.

Note that in the first instance,
the cross term, $\Sigma_{th} t h$, may
simply be dropped.  The resulting scaling contribution, $\Delta \Sigma
/ K |t|^\mu$, is plotted versus $\mt$, the scaled density deviation,
in Fig.~\ref{fig5:SMpm}.

Since the field-dependence of the background $\Sigma_0(t,h)$ has been
estimated only from data that lie on the two isotherms closest to
criticality and fall within the restricted range
$|\check h| \lsim 2.4 \negthinspace \times \negthinspace 10^{20}$,
we have distinguished, by solid dots \stress{versus} crosses and plusses,
between these observations, corresponding to the restricted range of $h$,
and (selected) further data falling outside this range.

Because the background contribution constitutes such a large part of the
total surface tension, it is not surprising that the subtraction needed
to isolate the singular piece, $\Delta \Sigma (t, h)$, is subject to
relatively large uncertainties.  This appears to be the main reason why
many of the crosses and plusses, derived from the low-density data
(beyond the fitted range of $h$), fall significantly below the scaling
loci.
For the data inside the range (the solid dots in Fig.~4),
the agreement between the
theoretical and experimental results is moderately encouraging considering the
uncertainties of the original data
and the inherent difficulties of the analysis.  Certainly, the data of 
{\bf NWW}
support the expected difference between $S_M^+(\mt)$
and $S_M^-(\mt)$.

It may be noted, nevertheless, that
just as pointed out by {\bf NWW},
the logarithmic singularity associated with the complete-wetting
transition cannot be detected in these or other surface-tension plots.
However, this is due not only
to the limited precision of the experimental data but also
to the fact that in the usual
representations of the data the singularity is not really visible
even in theoretical plots where it is known analytically to be present:
compare with Figs.~5 and 6 of {\bf I}.

If one also includes a \stress{positive} cross-coefficient
$\Sigma_{th}$ in the background
with the magnitude given in \whatis{eq5.4:S2x},
one finds slight changes of less than 1\% in
the data points plotted in Fig.~\ref{fig5:SMpm}.   However, when the
cross-term is inserted with a \stress{negative} sign, the majority
of the data points display changes of $\sim\negthinspace5$\% and a few
points prove still more sensitive.
This situation is hardly satisfying but, in light of the experimental
challenges and the resulting precision of the {\bf NWW} data,
it is clear that our analysis can proceed little further.

Finally, as pointed out theoretically by Ramos-G\'omez and Widom
\cite{Widom80} and mentioned briefly in {\bf I},
the surface tension isotherms when examined vs density
($\propto M$) above $T_c$ should, in general, exhibit
\stress{crossings}.
Although hints of such crossings have appeared in some binary fluid
systems \cite{Campbell68, McLure79, LastNote}, they were not seen in the
{\bf NWW} experiments.  The primary reason seems to be that for the
range of densities $\rho > \rho_c$ explored, the isotherms chosen
for observation above
$T_c$ were somewhat too widely spaced.  Indeed, as illustrated in
Fig.~\ref{fig6:isox}, the anticipated crossings for isobutyric acid $+$
water can be exhibited on the basis of our fits to the {\bf NWW} data.
All that is needed is to choose temperatures above $T_c$ spaced apart by
intervals of no more than $1 \thinspace \celc$ and to extend the
observations to densities slightly further above $\rho_c$.  The crossings,
while sensitive to the presence of the background term, should be
visible even in the singular part of the surface tension, $\Delta
\Sigma(T, \rho)$.  Note, however, that since our fits were based only on
the restricted (symmetric) range of $h$, the plots lying above the short
horizontal lines in Fig.~\ref{fig6:isox} may be less reliable than at
lower densities.

%
%
\section{Summary}
\noindent
In pioneering experiments to study critical endpoint behavior,
Nagarajan, Webb, and Widom \cite{NWW} made extensive measurements
of the vapor-liquid interfacial
tension of isobutyric acid and water mixtures as a function of
temperature at various compositions.  Since
the precision of the surface tension data was somewhat 
limited, we used the observations of the critical surface tension
made by Howland \etal~\cite{Knobler} and the coexistence curve
measurements of Greer~\cite{Greer} to cross-calibrate the {\bf NWW}
data.
We found full consistency between the various observations and verified
agreement with Antonow's rule near the critical endpoint.  In this
way reliable estimates of the coexistence curve amplitude $B$ and the
surface tension amplitude $K$ were obtained: see \whatis{eq5.3:amplB} 
and \whatis{eq5.3:Kexp}.
Using known universal amplitude ratios \cite{Zinn98}
and these two amplitudes, which can serve as the only required metric factors
in a hyperscaling formulation,
we estimated the nonuniversal susceptibility amplitude $C^+$.
Thereby the experimental data as a function of density could be
transformed in order to uncover the field-dependence of the crucial
surface-tension background contribution in the form (1.10).
Owing to the considerable uncertainties, however,
we could not resolve the cross-coefficient
$\Sigma_{th}$.  Nevertheless, even without this term, the scaling
plots for the singular part of the interfacial tension as derived from
the experimental data proved consistent with the theoretical
predictions based on the EdGF theory of {\bf I} \cite{Zinn03}:
see Fig.~\ref{fig5:SMpm}.

On the other hand, as discussed in the Appendix, significant
discrepancies come to light when the Howland \etal~data are compared,
via hyperscaling expectations, with evidence from other critical
systems.  Owing to conflicting experimental reports \cite{Kprob,
Moldover85}, however, it seems impossible to resolve this issue and
assess its significance without further experiments and, perhaps, new
theoretical insights.

More rewarding from a theoretical perspective and less open to questions
of precision, the critical surface tension
measurements of Mainzer-Althof and Woermann~\cite{Mainzer} on 2,6-dimethyl
pyridine $+$ water at the critical composition, show rather satisfactory
agreement
between the experimentally determined universal ratios $P$ and $Q$
[see (1.9)] and the EdGF predictions.  However, as regards the full
scaling aspects of the theory, more detailed, more precise and more
reliable data are sorely needed to execute more searching tests!

%
%
\begin{acknowledgments}
\noindent
We appreciate the keen interest of B.~Widom and comments from J.~Indekeu
and M.~R.~Moldover.
We are indebted to C.~M.~Knobler and I.~L.~Pegg for informative
discussions regarding their surface tension measurements.
The support of the National Science Foundation
through grants CHE 99-81772 and CHE 03-01101 is gratefully acknowledged.

Especially on this occasion, it is a pleasure for M.E.F. to acknowledge
the inspiration provided over more than four decades by the limpidly
insightful researches of Benjamin Widom.
\end{acknowledgments}

%
%
\appendix*
\section{Nonuniversal Critical Amplitudes}
\noindent
In Sec.~III the coexistence curve amplitude $B$ for isobutyric acid $+$
water (IBA:W) was determined from the data of Greer \cite{Greer} and
checked against the {\bf NWW} data yielding \whatis{eq5.3:amplB}.  Likewise,
analysis of the {\bf HWK} capillary-rise surface tension data yielded
the critical amplitude $K$ reported in \whatis{eq5.3:Kexp} that also
accords well with the {\bf NWW} data using forced capillary waves:
indeed, no larger deviation in \whatis{eq5.3:Kexp}, than, say, $\pm 6.0$ (or
$\pm 18$\%) could be reconciled with the {\bf NWW} observations
\cite{ThankCK}.  We stress this point because Moldover
\cite{Moldover85}, in comparing these IBA:W data with results for many
other fluid systems, uncovered an inconsistency in the value of the
expected-to-be universal amplitude ratio
\begin{equation}
S^+ \equiv \Kbar (\xi^+_1)^2 = 0.377 \pm 0.011
\quad\quad\quad \hbox{with} \quad\quad\quad
\Kbar \equiv K/k_B T_c. \label{eqA:Kxi2}
\end{equation}
Here, in the notation of the Appendix and Table~I of \cite{Zinn98},
the amplitude $\xi^+_1$ specifies the 
magnitude of the divergence of the second-moment
correlation length above $T_c$ via $\xi_1(T) \approx \xi^+_1 /t^\nu$,
while the value quoted for $S^+$ is that reported in \cite{Zinn98}.

If we now accept this theoretical result
we may proceed to predict the value of $\xi_1^+$
for IBA:W.  Thence, using the universal ratio \cite{Zinn98}
\begin{equation}
S_0 \equiv C^+ \Kbar / B^2 \xi^+_1 = 1.17 \pm 0.06,
\label{eqA:S0}
\end{equation}
together with the estimates for $K$ and $B$, we can compute the
susceptibility amplitude $C^+$ that is needed in Sec.~V to establish the
$h$ scale in Fig.~4.  This route, indeed, yields the {\bf A} value for
$C^+$ reported in \whatis{eq5.4:CpA} and, with the further aid of
\whatis{eq5.3:Scgl} and {\bf I} (6.9), the values for $K^c_>$ and $K^c_<$
given in \whatis{eq5.4:KcglA}.  These results, in turn, are incorporated in
the plots in Fig.~5 that serve to test the scaling hypothesis.

The correlation length amplitude predicted in this fashion is $\xi_{\rm
1, pre}^+ =
2.10 \pm 0.08 \thinspace \Ao$ [where the {\bf HWK} value
\whatis{eq5.3:KnoblerTc}
has been adopted for $T_c$ in \whatis{eqA:Kxi2}].  However, as reported by
Moldover \cite{Moldover85},
in light scattering experiments by Chu and co-workers~\cite{Chu73, Chu68},
the second-moment
correlation length above $T_c$ was evaluated for IBA:W.
The fits presented led them to conclude
$\nu \simeq 0.613$ and $\xi_1^+ = 3.57 \pm 0.07 \thinspace \Ao$.
This correlation amplitude
exceeds the predicted value, $\xi^+_{1, {\rm pre}}$, by 60\% or more!
On re-analyzing their
data with the higher value $\nu = 0.632$ imposed, we obtain
$\xi_1^+ = 3.09 \pm 0.05 \Ao$.  This is somewhat closer to
$\xi_{1, \rm pre}^+$
although the uncertainty limits fail to overlap by 36\%.
From turbidity
measurements, Beysens \etal~\cite{Beysens} concluded $\xi_1^+ = 3.62_5 \pm
0.06_5 \Ao$
and $\nu \simeq 0.630$; this value of $\xi_1^+$ agrees with Chu
\etal~and is some 17\% higher than that from
our reanalysis.

At a basic theoretical level the discrepancies uncovered here are of
serious concern.  If, for concreteness, one accepts
$\xi^+_{\rm 1, expt} \simeq 3.6_0 \pm 0.07 \thinspace \Ao$
as the observed value, in accord with these experiments,
the universal ratio $S^+$ in \whatis{eqA:Kxi2}
would yield a prediction for $\check K$ of only 12.0,
about 36\% of that observed!  Conversely, one might be tempted to assert
that the
experimentally determined value of the (supposedly universal) ratio
for IBA:W is $S^+_{\rm expt} = 1.06 \pm 0.03$, in place of
\whatis{eqA:Kxi2},
the uncertainty indicated here taking into account various sources of
experimental and theoretical imprecision.

Similarly, the {\bf B} values for $C^+$ and $K^c_{\gl}$ reported in
\whatis{eq5.4:CpB} and \whatis{eq5.4:KcglB} follow from the observed value of
$B$ and the experimental value, $\xi_{1, \rm expt}^+$, by using
\cite{Zinn98}
\begin{equation}
Q_c \equiv (\xi_1^+)^3B^2/C^+ = 0.323_6 \pm 0.006.
\end{equation}

In facing this discrepant situation let us focus first on the
theoretical value \whatis{eqA:Kxi2} for $S^+$ (which, indeed, was
incorporated into the numerical scaling calculations reported in {\bf I}
and tested here).  One might initially notice that 0.377 is
significantly higher than the various theoretical, RG and Monte Carlo
estimates, namely, $\simeq \negthinspace 0.21$, $0.24$, and $0.28$,
available to Moldover in 1985 \cite{Moldover85}.
In fact \whatis{eqA:Kxi2} rests on the analysis \cite{Zinn} of difficult but
carefully performed Monte Carlo simulations for the nearest-neighbor
$(d=3)$-dimensional Ising model by Hasenbush and Pinn \cite{Pinn94}
(combined with the well established universal ratio $\xi_1^+ / \xi_1^- =
1.96 \pm 0.01$ \cite{Zinn, Liu}).  
The simulations \cite{Pinn94} could
approach criticality no closer than $|t| > 0.015$.  By contrast, the
experimental data for IBA:W
lie in the region $|t| = 0.0008$ to $0.010$.  Although
it seems unlikely, it is conceivable that
exceptionally strong correction-to-scaling terms in the surface tension
of the simple cubic Ising model could dominate very close to $T_c$
so that \whatis{eqA:Kxi2} represents a significant under-estimate of $S^+$.
Alternatively, finite-size effects in the simulations may be playing a
far larger role than appreciated although no evidence to suggest this
was observed and the issue was by no means neglected \cite{Gelfand90,
Pinn94}.

On the other hand, it might be that for reasons associated with the
lattice structure and capillary-wave fluctuations, the standard simple
cubic lattice gas/Ising model is inadequate for certain real fluid
mixtures.  Most certainly, the conventional near-neighbor
lattice gases neglect the long-range power-law van der Waals
interactions that lead to interphase potentials decaying, normal to
an interface, only as $1/z^3$.  And while these van
der Waals forces play only a \textit{subdominant} role in \textit{bulk} critical
behavior, they are known to be especially significant in surface and
wetting phenomena \cite{Dietrich}.  It is, thus, possible that
they seriously distort the near critical interfacial tension or even
destroy the expected universal character of the ratio $S^+$.

At present, however, effective theoretical tools for addressing this
interesting issue are not apparent.  Furthermore, it is not unfair to
remark that these various rather deep theoretical issues do not directly
affect our
basic \textit{scaling} \textit{analysis} of the {\bf NWW} observations
of surface tensions near a critical endpoint.  Indeed, in the first
instance only the experimentally determined critical amplitudes $K$ and
$B$ enter: universal or nonuniversal relations between these and other
critical amplitude do not play a definitive role in the primary scaling
issues.

Finally, we must return to the experimental evidence.  It is, indeed,
striking that the value for $S^+$ in \whatis{eqA:Kxi2} accords remarkably
well with the experimental values assembled by Moldover
\cite{Moldover85} for Ar, Xe, ${\rm N}_2$, ${\rm O}_2$, ${\rm CO}_2$,
${\rm CH}_4$ and ${\rm SF}_6$ and for the
binary fluids triethylamine $+$ water, nitroethane $+$ 3-methylpentane,
cyclohexane $+$ aniline,
methanol $+$ cyclohexane, and even for various polystyrene $+$ ${\rm
C}_7{\rm H}_{14}$ solutions: see Fig.~1, Table I and also the {\it Note
added\/} regarding Ref.~86 in \cite{Moldover85} and Table I of
\cite{Moldover86}.  Indeed, by and large, Moldover's analysis confirms
rather generally the hyperscaling hypothesis, \stress{both} for the
interfacial \stress{and} for the bulk thermodynamic properties. (See also
in \cite{Zinn1997}.)

These observations tend to reinforce confidence in the estimate
\whatis{eqA:Kxi2} for the universal surface tension amplitude---at least as
regards `simple' pure fluids and binary mixtures.  But does the weak
acid IBA:W
constitute such a `simple' system?  On the one hand, as an electrolyte,
a population of positive and negative ions will be present in all the
liquid phases.  Furthermore, it is known that the critical point of
IBA:W is
extremely sensitive to small impurities and, especially, to
ionic impurities \cite{Greer, Knobler05}.  Ionic impurities, in
particular, may be expected to have significant effects on the various
liquid-liquid, liquid-vapor and liquid-solid interfaces; but even in the
absence of impurities, an interphase Galvani potential difference will
appear across the critical liquid-liquid interface together with an
associated electrically charged double layer \cite{REF51, REF52, REF53,
REF54}.

On the other hand, Moldover \cite{Moldover85} in addressing
the discrepancy issue
notes, as a private communication from I. L. Pegg and C. M. Knobler,
that further surface-tension measurements on IBA:W using a different
technique (actually light scattering from thermally excited capillary
waves \cite{Knobler05}) yielded the preliminary result $\check K = 11.6
\pm 1.0$.  This value is almost three times smaller than found in the
{\bf HWK} experiments with which, as indicated above, it cannot be
reconciled.  Nevertheless, if this value for $\check K$ is
used with the experimentally determined values,
$\xi_{1, \rm expt}^+$, discussed above it yields a value for $S^+
\propto \check K (\xi_1^+)^2$ close to that found experimentally for the
other systems and, hence, in \textit{accord} with the expected universal
value stated in (A.1) \cite{Moldover85, Moldover86}!  On this basis one
might (despite \cite{ThankCK}) be tempted to disregard the {\bf HWK} data
and argue that our analysis of the {\bf NWW} observations should be
seriously reconsidered.

Unfortunately, the preliminary study of IBA:W by Pegg and Knobler has
never been published \cite{Knobler05}; nor, to our knowledge, have any
subsequent measurements been made by others on this system.
Thus, while one can and,
perhaps, should attribute the nearly
three-fold difference in the critical
surface tension to unidentified impurities or other agencies
distinguishing the {\bf HWK}
and {\bf NWW} samples of isobutyric acid from the later sample
investigated by Pegg and Knobler (in 1984), the true causes
of the discrepancies and their
significance remain obscure.  We must, instead, hope that more sensitive
techniques will be developed and employed on the IBA:W and 
other binary systems in
order to more stringently test the scaling theories for critical
endpoints that have been developed.

%
%
\bibliography{ms}%

\begin{thebibliography}{54}
\bibitem[{Gre({\natexlab{b}})}]{GreerDiameter}
\bibinfo{note}{In contrast to the {\bf NWW} data, the coexistence curve
  measured by Greer~\cite{Greer} is somewhat less symmetric: on the basis of
  her data, the diameter can be represented by ${\check{\overline \rho}}(T)
  \simeq 0.993 - 0.066 t$.}
\bibitem[{Las()}]{LastNote}
\bibinfo{note}{We are indebted to B. Widom for drawing our attention to Refs.
  \cite{Campbell68} and \cite{McLure79}.}
\end{thebibliography}
\centerline{\vbox{\hrule width 3 in height 2 pt}}

%
%
\newpage
\begin{table}
\caption{Adopted values of fluid critical exponents.  For convenience
and consistency with the universal amplitude ratios assembled in
\cite{Zinn98} we have employed the corresponding exponent values
(satisfying the scaling relations).  More recent studies
\cite{Butera2000, Pelissetto2002, Guida1998, Campostrini2002}
lead to our currently
preferred exponent estimates shown here in parentheses.  However, for
the present analysis the differences are of negligible consequence.
We may supplement the relations \whatis{eq5.1:exprel} with $\Delta =
\beta \delta = \beta + \gamma$ and $\gamma = (2 - \eta) \nu$.}
\label{tbl:exp}
\begin{ruledtabular}
\begin{tabular}{c c c c c}
\emspace\emspace\emspace $\beta$ & $\gamma$ & $\nu$ & $\mu$ &
	$\theta$ \emspace\emspace\emspace\\
\hline
\emspace\emspace\emspace 0.3266 & 1.2392 & 0.6308 & 1.2616 &
	0.54 \emspace\emspace\emspace\\
\emspace\emspace\emspace (0.3260) & (1.2390) & (0.6303) & (1.2606) &
	(0.52) \emspace\emspace\emspace\\
\end{tabular}
\end{ruledtabular}
\end{table}

%
%
\hbox to 1 in { }
\newpage
\noindent
{\bf FIGURE CAPTIONS}

\noindent
FIG.~1.  
The coexistence curve of the isobutyric acid and water
mixture.  The dots represent the experimental data of {\bf NWW}.
The inner curve is the fit \whatis{eq5.3:form} using the
amplitude $\check B = 0.0310$ with $a_\theta$ and $a_1$
from \whatis{eq5.3:Brho} while the outer curve employs $\check B = 0.0318$
in accord with \whatis{eq5.3:Brho2}; the critical temperature has been taken
as $T_c = 26.310 \celc$ in accord with the estimate of {\bf NWW}.
\bigskip

\noindent
FIG.~2.  
Comparison of surface-tension measurements along the coexistence
curve: the open circles with error bars represent the {\bf NWW}
data. As explained in the text, the parallelograms have been drawn
using $\Sbr = K|t|^\mu$ with the amplitude
\whatis{eq5.3:Kexp} obtained from the {\bf HWK} data.
The vertical sides of the parallelograms indicate the ranges in which
consistency between the two experiments can be achieved
in accord with Antonow's rule.  The dashed curve,
is based on the fit \whatis{eq5.4:h0fit} obtained below for the background
$\Sigma_0(T,0)$,
while the solid curve embodies the values \whatis{eq5.4:h0fitb}.
See text in Sec.~IV.A for details.
\bigskip

\noindent
FIG.~3.  
The fit to the surface tension background in zero-field.  See text
for details of the generation of the data points (black dots).  The dashed
curve is drawn using the estimates \whatis{eq5.4:h0fit}.   The solid curve
represents the fit \whatis{eq5.4:h0fitb}.
\bigskip

\noindent
FIG.~4.  
A fit for the surface tension on the critical isotherm
as a function of the ordering field.  The dots
joined by the dashed lines represent the {\bf NWW}
isotherms nearest the critical
temperature that serve as approximate
upper and lower bounds.  See text for details.
\bigskip

\noindent
FIG.~5.  
Comparison of scaling plots
for the vapor-liquid surface tension near the critical endpoint of
isobutyric acid $+$ water.  The solid curves represent the theoretically
predicted scaling functions $S_M^\pm(\mt)$ using the ameliorated EdGF theory.
The dots are derived from the {\bf NWW}
data for $T>T_c$ and $T<T_c$ that fall within the symmetric
range $|\check h| \lsim 2.4 \negthinspace \times \negthinspace 10^{20}$
used in fitting the surface tension background.
The other symbols ($\times$ for $T>T_c$ and $+$ for $T<T_c$)
are some of the {\bf NWW} data
falling outside the symmetric range of $h$ (and hence not covered by the
background fit).
Note that an offset of $\Delta S = 100$
has been used for clarity in the vertical scales
for $T \gl T_c$.  See the text for further details.
\bigskip

\noindent
FIG.~6.  
Illustration of predicted crossings of surface-tension isotherms of
isobutyric acid $+$ water above $T_c$.
The isotherms have been computed using the EdGF scaling functions
$S^\pm_M(\mt)$ from {\bf I} with a background of the form~\whatis{eq5.1:S0}
suitably matched to the experimental data.
The dotted, solid, dashed, and dot-dashed curves represent isotherms
$\Delta T \equiv T - T_c = 0$,
$1.0$, $2.0$, and $3.0 \celc$, respectively.
The narrow horizontal and vertical lines mark
the critical values of $\Sigma$ and $\rho$.
The short horizontal lines mark the ordering field values up to which
the fitting was performed.
See the text for further details.

%
%

\begin{figure}
\includegraphics[scale=0.95]{fig2_chk}
\caption{\label{fig5:chk}
}
\end{figure}

\begin{figure}
\includegraphics{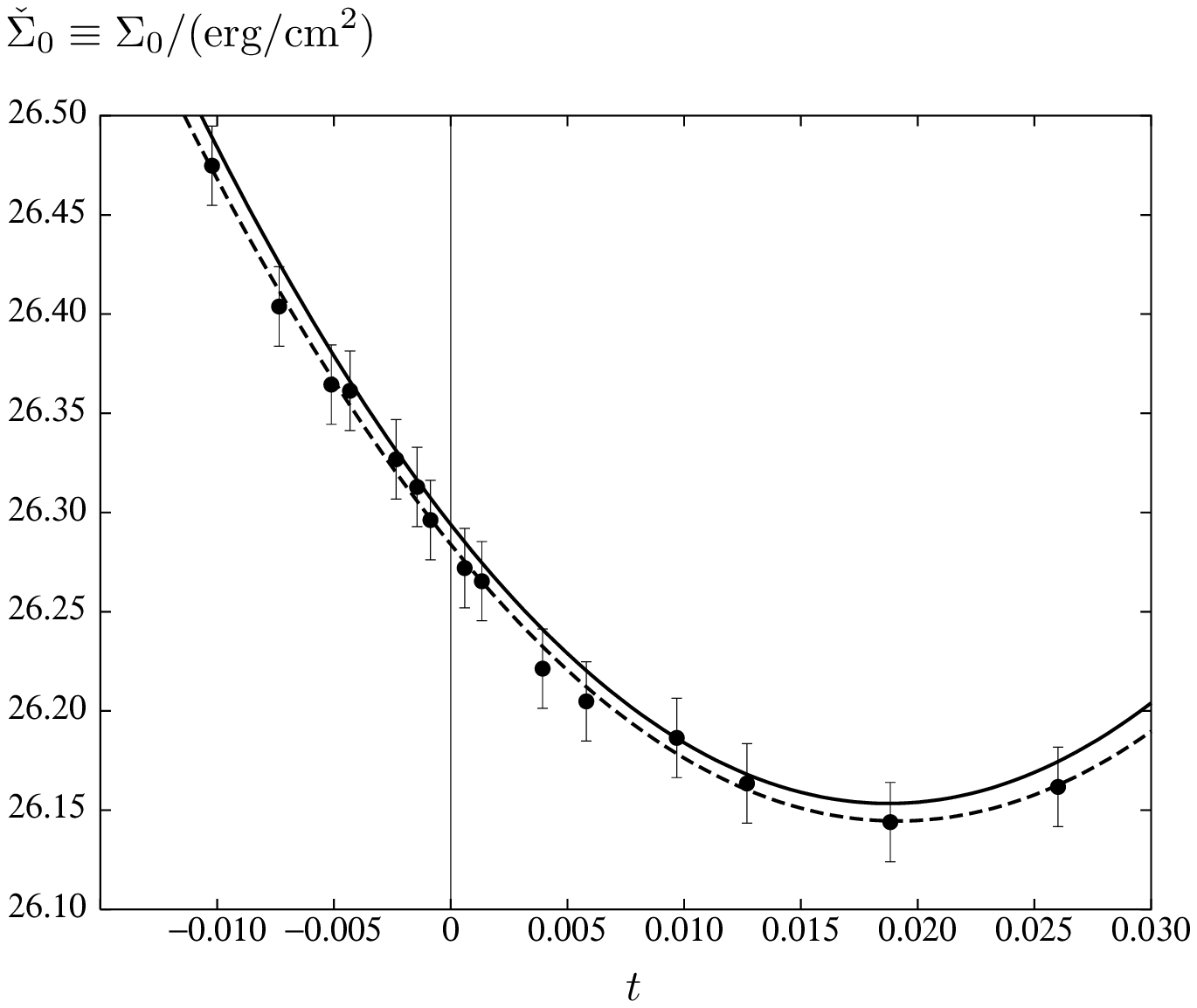}
\caption{\label{fig5:sigma0}
}
\end{figure}

\begin{figure}
\includegraphics{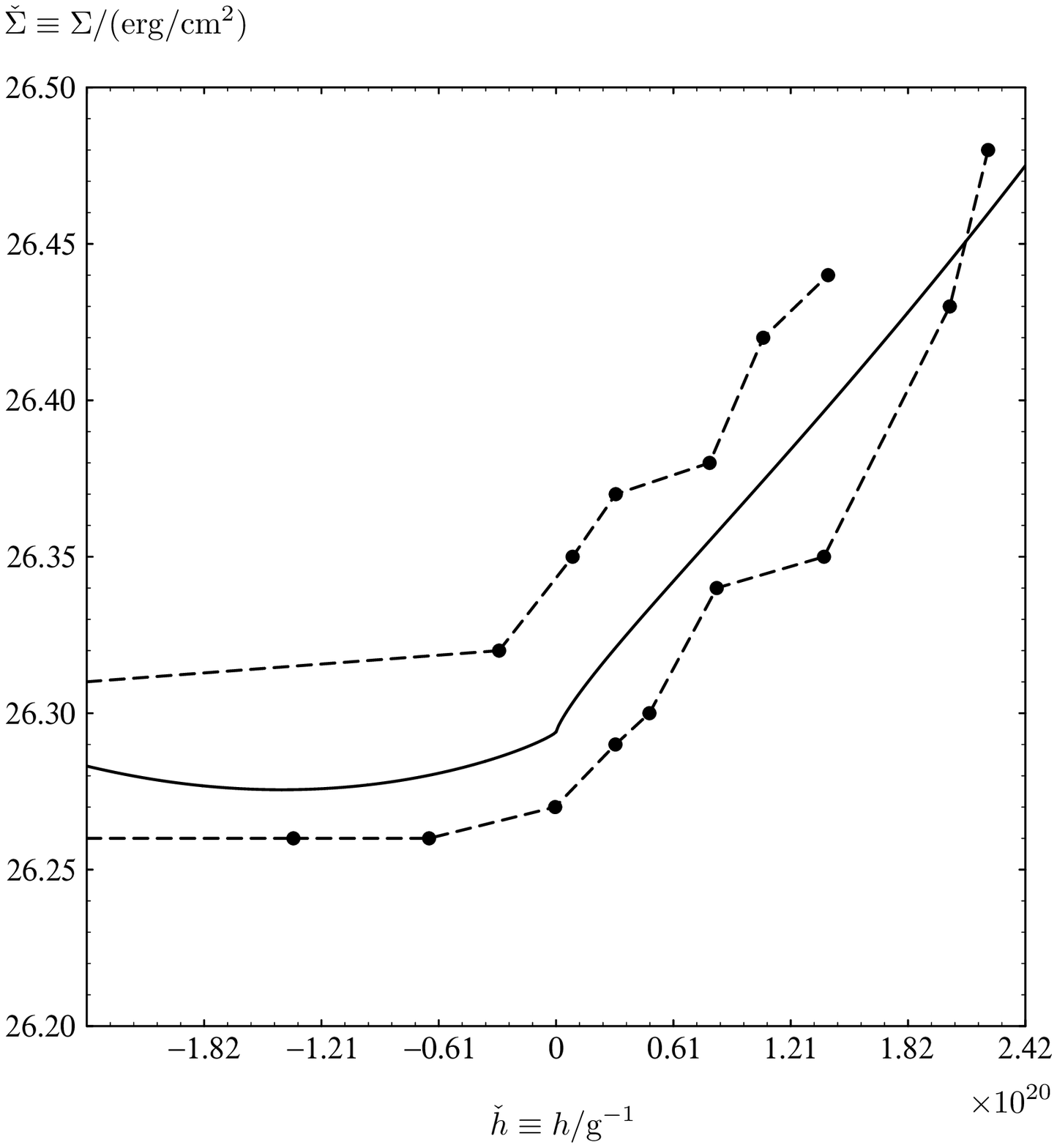}
\caption{\label{fig5:attc}
}
\end{figure}

\begin{figure}
\includegraphics[scale=0.85]{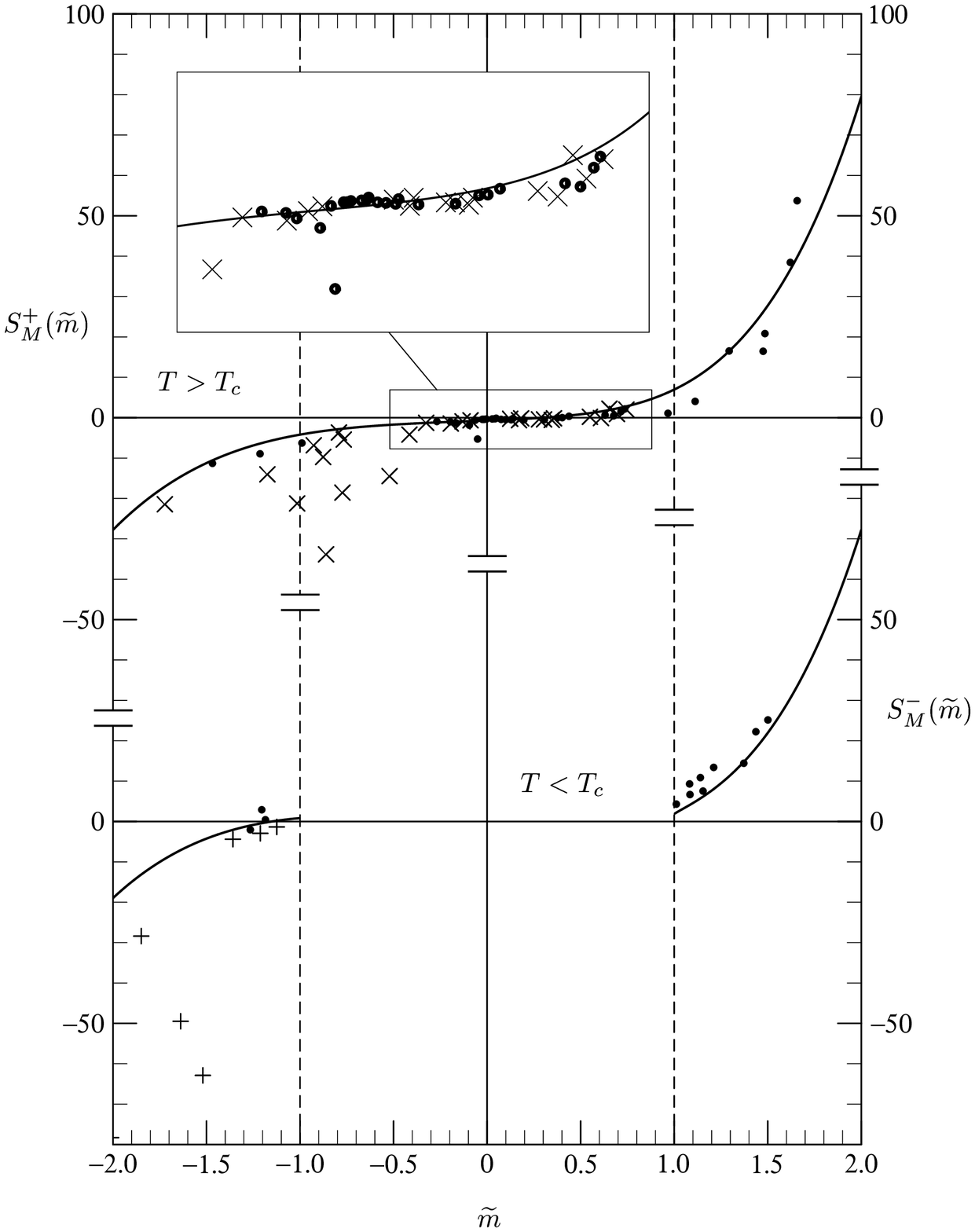}
\caption{\label{fig5:SMpm}
}
\end{figure}

\begin{figure}
\includegraphics[scale=1.0]{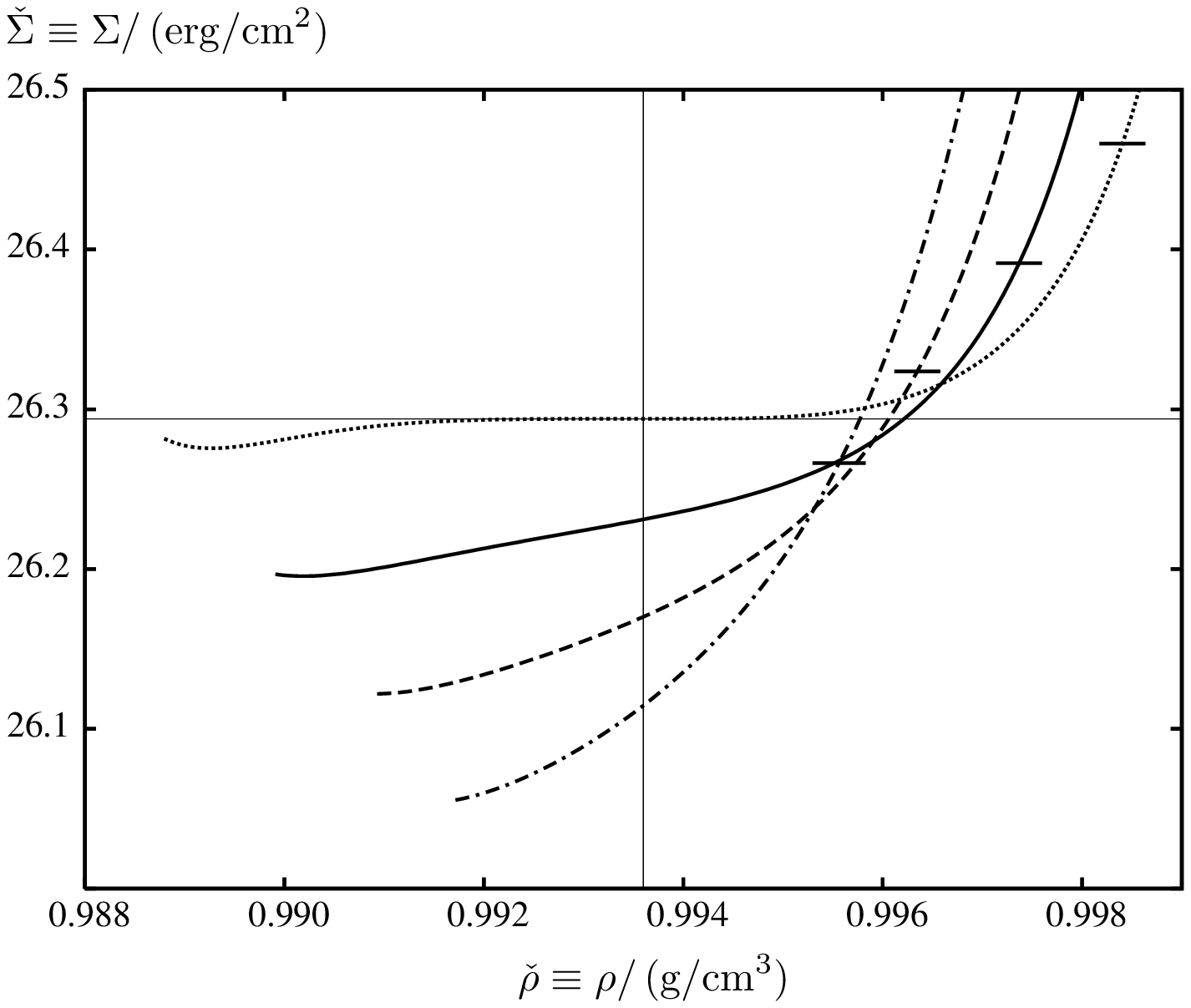}
\caption{\label{fig6:isox}
}
\end{figure}

\end{document}